\documentclass{aa}
\usepackage{graphicx,epsfig}

\begin{document}
   \title{Spin orientation of supermassive black holes in active galaxies}

   \author{W. Kollatschny
         \fnmsep
          \thanks{Based on observations obtained with the Hobby-Eberly
               Telescope, which is a joint project of the University
               of Texas at Austin, the Pennsylvania State
           University, Stanford University, Ludwig-Maximilians-Universit\"at
            M\"unchen, and Georg-August-Universit\"at G\"ottingen.}
          }

   \offprints{W. Kollatschny}

   \institute{Universit\"{a}ts-Sternwarte G\"{o}ttingen,
              Geismarlandstra{\ss }e 11, D-37083 G\"{o}ttingen, Germany\\
              \email{wkollat@uni-sw.gwdg.de}}

   \date{Received date, 2003; accepted date, 2003}

   \abstract{
Accretion of gas onto a central supermassive black hole is
generally accepted to be the source of the emitted energy in active galactic
nuclei.  The broad emission lines we observe in their optical spectra are
probably formed in the wind of an accretion disk at distances of light days to
light years from the central black hole. The variable fraction of the
emission lines originates at typical distances of only 1 to 50 light days
from the central supermassive black hole. We derived a central black hole
mass of  $M_{\rm orbital}=1.8\pm 0.4\times 10^{7} M_{\odot}$  in the Seyfert
 galaxy Mrk\,110 assuming
the broad emission lines are generated in gas clouds orbiting within an
accretion disk. This figure depends on the inclination angle of the accretion
disk. Here we report on the detection of gravitational redshifted emission
in the variable fraction of the broad emission lines. We derive a central
black hole mass of $M_{\rm grav}=14.0\pm 3.0\times 10^{7} M_{\odot}$. These
measurements are
independent on the orientation of the accretion disk. The comparison
of both black hole mass estimates allows to determine the projection
of the central accretion disk angle $i$ to $21\pm5 \deg$. in Mrk\,110 and
therefore the orientation of the spin axis of the central  black hole. 
   }
  
  \authorrunning{W. Kollatschny}
  \titlerunning{Spin orientation of supermassive black holes}
  \maketitle
%

\section{Introduction}
Mass and spin of the central black holes in galaxies as well as geometry and
kinematics of their immediate stellar and gaseous surroundings are of
fundamental interest for understanding the origin of the enormous energies
emitted in the nuclear regions of active galaxies (e.g. Blandford \& Begelman
\cite{blandford99}, Murray \& Cheng \cite{murray97}). Furthermore, the central
black hole masses and their evolution might be connected to the general
properties of their host galaxies. The black hole mass has been estimated
in some active galaxies by reverberation mapping (e.g.
Wandel et al. \cite{wandel99}).
A basic assumption of
this method is that the broad emission line gas is gravitationally dominated
by the central black hole. Here a major uncertainty is introduced by
the geometry and orientation of the line-emitting gas.  An independent
mass determination of the central black hole can be obtained by the
gravitational redshift effect in the optical emission lines of active
galaxies.  However, in this respect only the detection of a gravitationally
redshifted Fe K$\alpha$ X-ray line has been reported in a few active 
galaxies (Tanaka et al. \cite{tanaka95}, Fabian et al. \cite{fabian00}). 

\section{Observations and data reduction}

We selected the narrow-line Seyfert 1 galaxy Mrk\,110 as our primary
target to determine the central black hole mass independently both by
reverberation mapping and by gravitational redshift effects. From earlier
long term campaigns we knew of the extreme continuum and line intensity
variations in this galaxy
(Bischoff \& Kollatschny \cite{bischoff99}, Peterson et al. \cite{peterson98}).
We carried out  our optical variability campaign with the 9.2m Hobby-Eberly
Telescope (HET) at McDonald Observatory. We took 26 spectra of Mrk\,110 between
1999 November 13 and 2000 May 14. All observations were made under identical
conditions with exactly the same instrumentation at the HET. The spectra
cover a wavelength range from 4200\AA\ to 6900\AA\ with a resolving power
of 650. 
The data were reduced in a homogeneous way with IRAF reduction packages.
In most cases we reached a
S/N $>$ 100 per pixel in the continuum.
Great care was taken of a very accurate internal calibration of the spectra
with respect to wavelength and flux
(Kollatschny et al.\cite{kollatschny01}).

\section{Results and discussion}

\subsection{Mean and rms line profiles}

We determined mean and rms emission line profiles of the most intense lines
H$\alpha$,
H$\beta$, HeI$\lambda$5876, and  HeII$\lambda$4686 from
our variability campaign.
The rms profiles are a measure of the variable component of the line profile.
The profiles shown in Fig.~1 are plotted in velocity space and normalized
to the same maximum  intensity.
\begin{figure*}
 \hbox{\includegraphics[bb=40 90 380 700,width=55mm,height=85mm,angle=270]{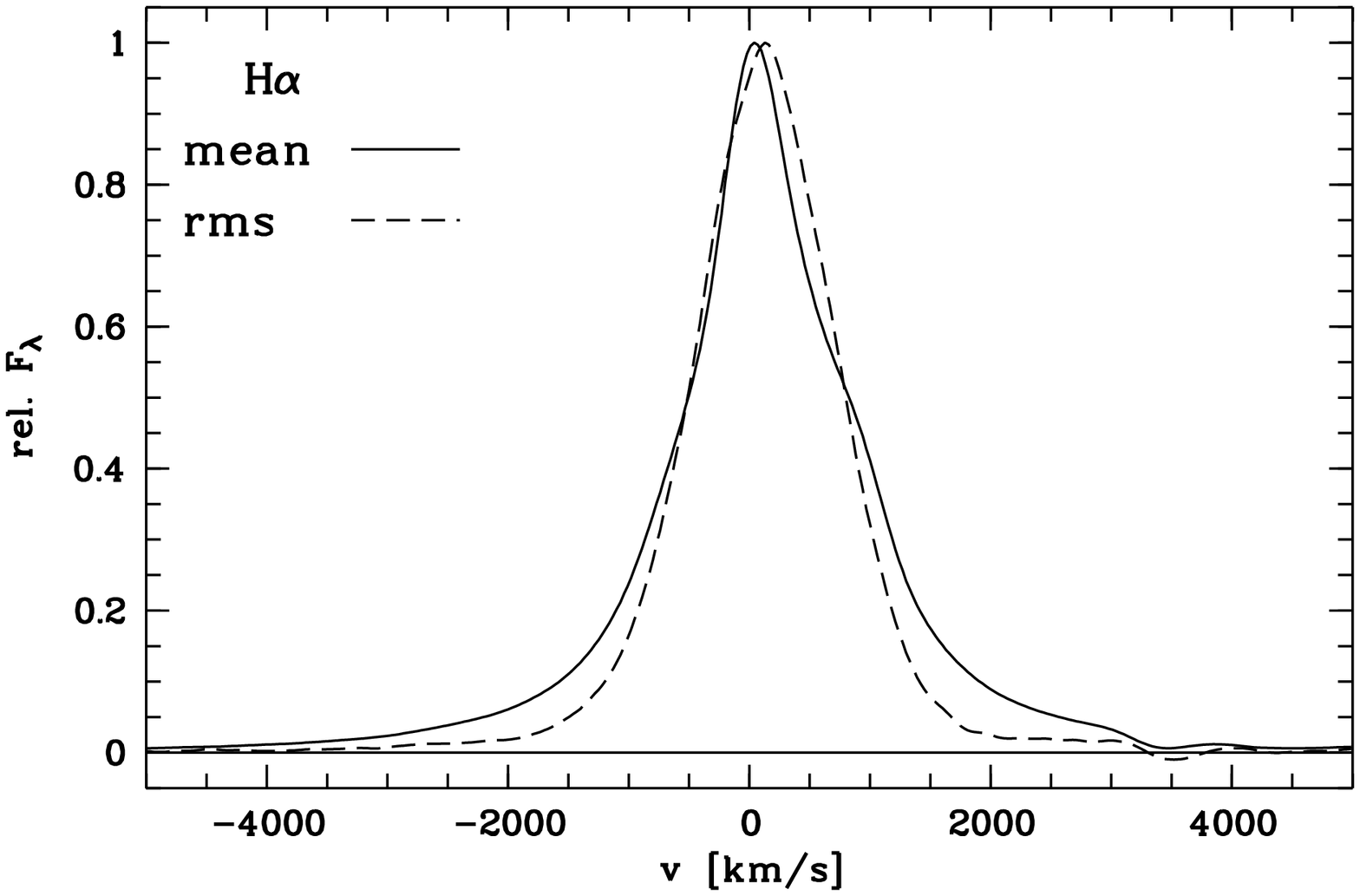}\hspace*{7mm}
       \includegraphics[bb=40 90 380 700,width=55mm,height=85mm,angle=270]{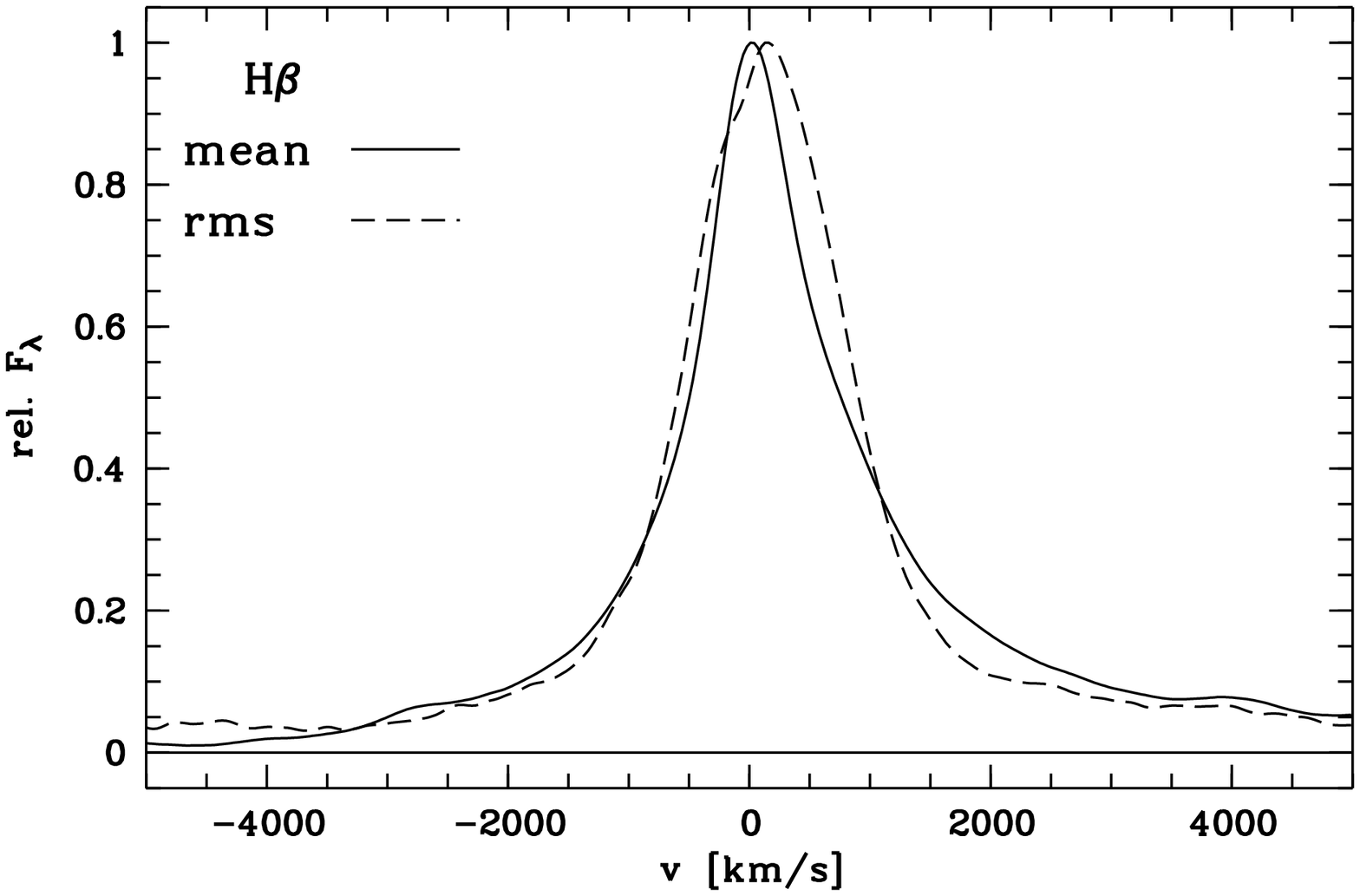}}
 \hbox{\includegraphics[bb=40 90 380 700,width=55mm,height=85mm,angle=270]{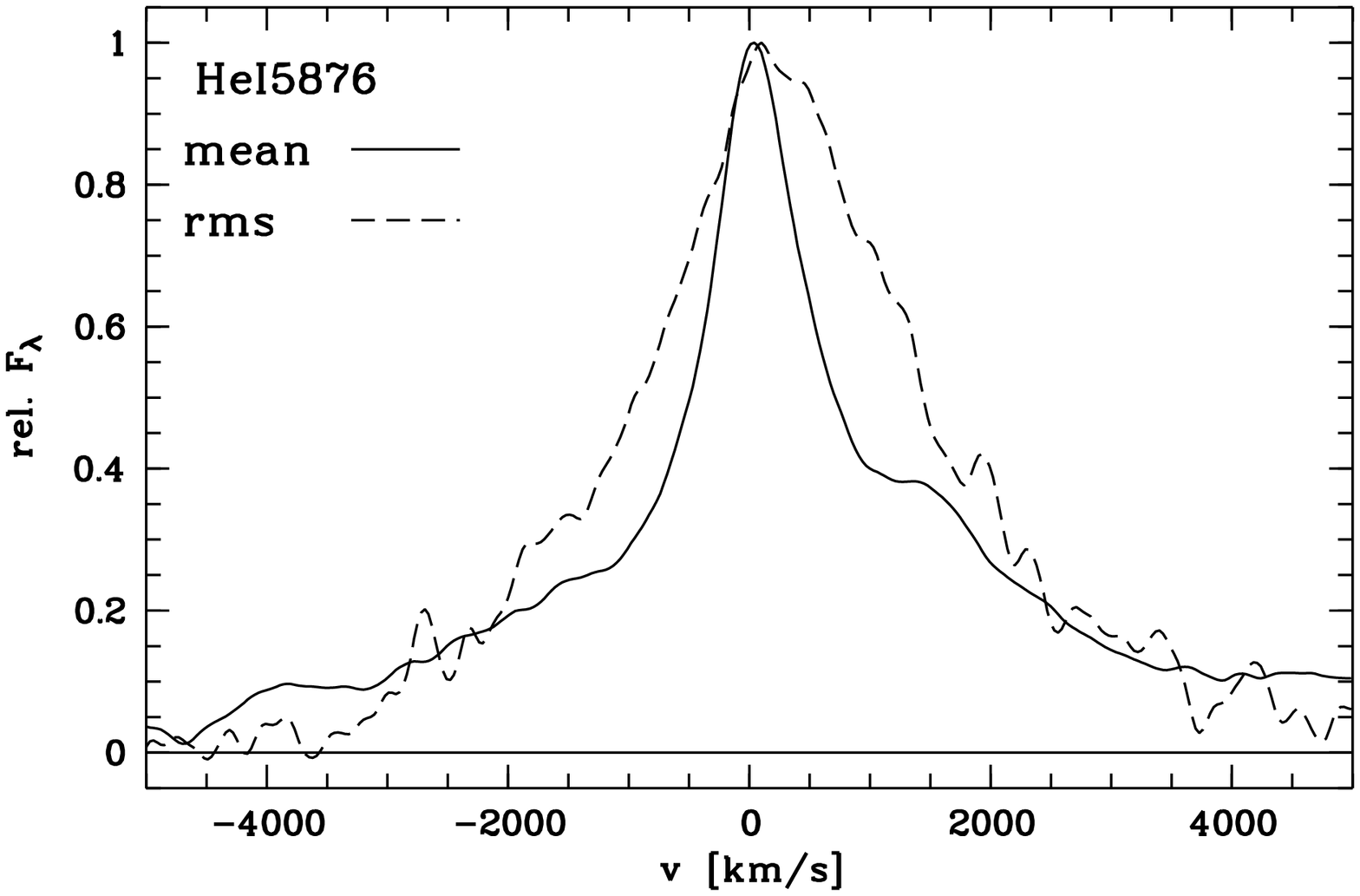}\hspace*{7mm}
       \includegraphics[bb=40 90 380 700,width=55mm,height=85mm,angle=270]{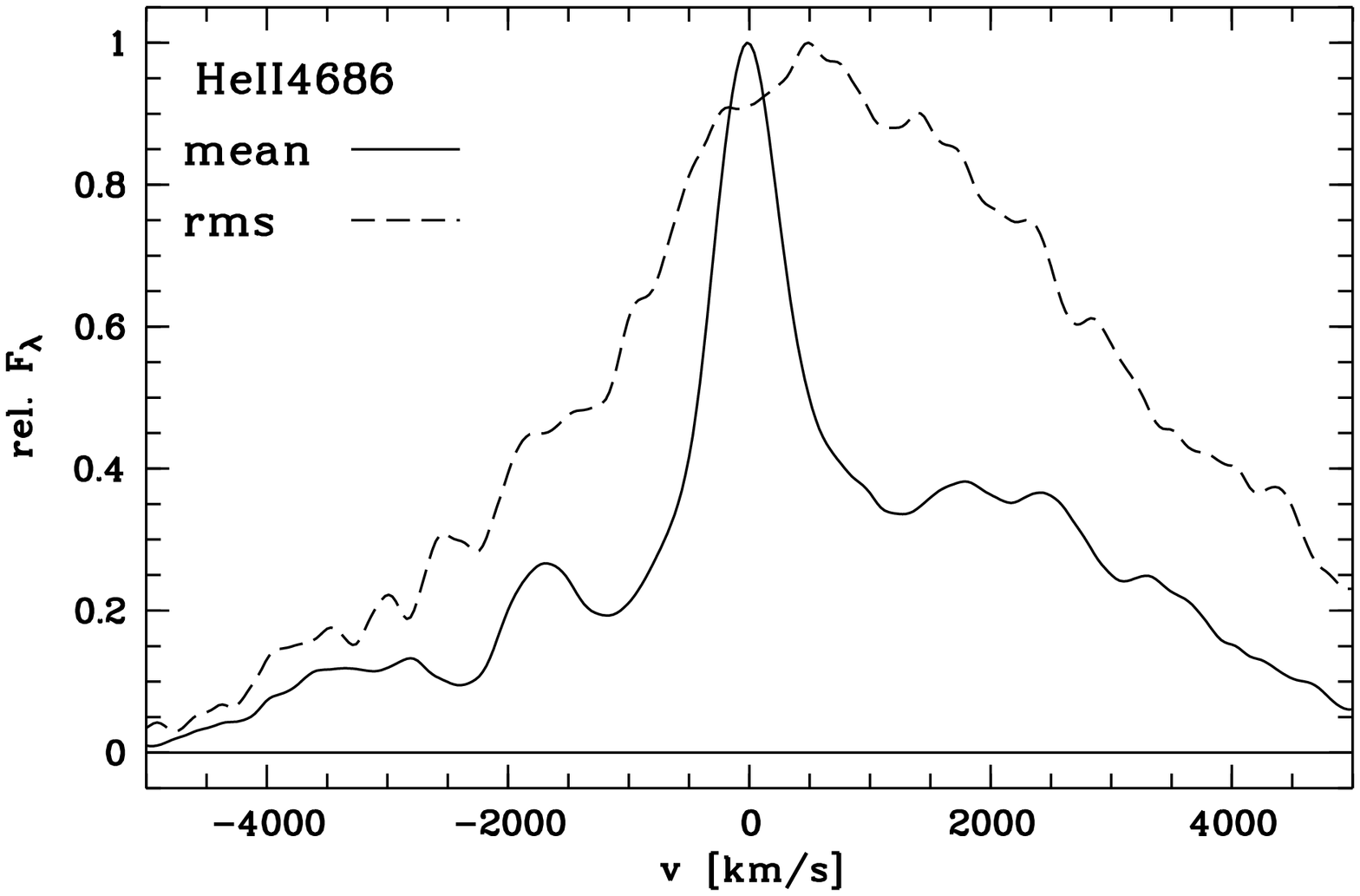}}
       \vspace*{5mm} 
  \caption{Normalized mean (dashed lines) and rms (solid lines)
           Balmer and Helium emission line profiles.}
\end{figure*}
The fraction of the variable component in these lines amounts to 9, 12, 14,
and 39\% , respectively. The line widths (FWHM) of the rms profiles
are listed in Table~1.

\begin{table}
\caption{Rms line widths (FWHM) of our strongest emission lines:
 H$\alpha$, H$\beta$, HeI$\lambda$5876, and  HeII$\lambda$4686; differential
 redshift of rms line centers $\Delta$\,$v_{\rm cent}$(rms) ; cross-correlation
lags $\tau$ and central black
hole mass estimation $M_{\rm grav}$ derived from gravitational redshift.}
\begin{tabular}{lcccc}
\hline
\noalign{\smallskip}
Line & FWHM(rms) & $\Delta$\,$v_{\rm cent}$(rms) & $\tau$ & $M_{\rm grav}$\\
     & [km s$^{-1}$] & [km s$^{-1}$] & [days] & [$10^7 M_{\odot}$]\\
(1) & (2) & (3) & (4) & (5)\\
\noalign{\smallskip}
\hline
\noalign{\smallskip}
HeII~ & 4444 $\pm$ 200 & 541 $\pm$ 60 & 3.9 $\pm$ 2 & 13 $\pm$ 3\\
HeI~ & 2404 $\pm$ 100 & 186 $\pm$ 60 & 10.7 $\pm$ 6 & 12 $\pm$ 4\\
H$\beta$ & 1515 $\pm$ 100 & 118 $\pm$ 50 & 24.2 $\pm$ 4 & 17 $\pm$ 4 \\
H$\alpha$ & 1315 $\pm$ 100 &  74 $\pm$ 50 & 32.3 $\pm$ 5 & 14 $\pm$ 5 \\
\noalign{\smallskip}
\hline
\end{tabular}
\end{table}
\subsection{Virial mass of the central black hole}

We calculated the characteristic distances of the line emitting regions by
cross-correlating the light curves of the emission lines with the
optical continuum light curve. The broader emission lines originate
systematically closer to the galaxy center in Mrk\,110 (see Table~1)
(Kollatschny et al. \cite{kollatschny01}, Kollatschny \cite{kollatschny03}).
The fraction of the variable line component  is correlated with the
distance of the line emitting region from the black hole. The distances
of the line emitting regions we determined from the cross-correlation
function (CCF) analysis correspond to the widths of the rms profiles 
(Table~1). The non-variable components of the emission lines originate
at larger distances from the black hole. We determined the central black
hole mass assuming Keplerian orbits of the line emitting clouds:
\[ M_{\rm orbital} = f v^{2} G^{-1} R .\]
where v is the characteristic emission-line velocity (rms line width), $G$ is
the gravitational constant, and $R = \tau$c is the distance of the
line emitting
region ($\tau$ = calculated cross-correlation lag of line emitting region).
The derived black hole mass in Mrk\,110 is
\[ M_{\rm orbital}= 1.8\pm0.4\times 10^{7} M_{\odot}. \]

The dimensionless factor $f$ depends amongst other things on the geometry
and (unknown) orientation of the BLR. We could show that the velocities
in the broad emission line region are not randomly oriented as for
instance in a turbulent flow but that the lines are formed in 
an accretion disk (Kollatschny et al. \cite{kollatschny02}, Kollatschny
\cite{kollatschny03}). The line widths of the rms spectral lines are
interpreted as the line-of-sight Doppler widths. We observe only
the projected velocity $v \sin i$ of the rotating accretion disk. Thus,
the derived mass is a lower limit. The systematic errors in the estimation
of black hole masses by reverberation mapping have been discussed by
Krolik (\cite{krolik01}). He pointed at a systematic underestimate of the
central AGN mass caused by the possible inclination of the line
emitting region.
\subsection{Emission line shifts}
 We determined from the spectra in
Fig.~1 the shift of the rms line centers (uppermost 20\%).
It is a known effect that rms line profiles sometimes
 show narrow forbidden line
residuals that are caused by variable seeing and small shifts in wavelength
calibration (e.g. Kaspi et al. \cite{kaspi00}).
 The degree of this contamination 
is strongest in the HeI$\lambda$5876 line which is the weakest one.
The redshift ($\Delta v$) of the rms profiles with respect to the
narrow emission lines increases as a function of line width and ionization
potential as seen in Fig.~1. The correlation of rms line width with
observed central line shift is shown in Fig.~2.
\begin{figure}
\includegraphics[width=100mm,height=98mm,angle=270]{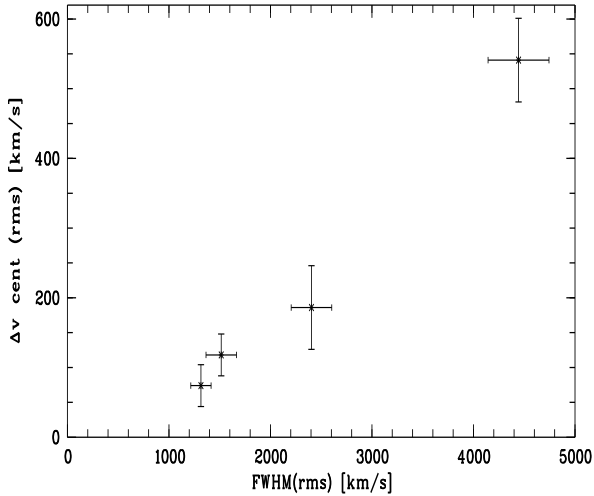}
\vspace{-28mm}
\caption{Relative redshift of H$\alpha$, H$\beta$, HeI$\lambda 5876$, and
HeII$\lambda 4686$  rms line centers as a function of rms line width
(FWHM)(see Table~1).}
\end{figure}

High-ionization broad emission lines originate closer to the central
supermassive black hole than the low-ionization ones (e.g. Peterson \& Wandel
 \cite{peterson99}, Kollatschny et al. \cite{kollatschny01}).
The redshifts derived from different integrated emission lines in
AGN often do not agree with each other. High-ionization broad emission
lines are sometimes blueshifted with respect to the low-ionization broad
emission lines and narrow emission lines (Gaskell
\cite{gaskell82}, Sulentic et al. \cite{sulentic00}).
Therefore, the shift of
the integrated broad line profiles cannot be explained by
gravitationally redshift
effects. But is has been demonstrated recently that the observed blueshift
in the highly ionized lines is mimicked by absorption in the red
line wings (Richards et al. \cite{richards02}). This effect
is more pronounced in the higher
ionized lines. Additionally, one has to consider that the integrated
emission lines originate at distances of light weeks to light years
from the black hole. A verification of a gravitational redshift
in these lines is beyond the detection limit.
\subsection{Central black hole mass derived from gravitational redshift}
The variable
broad line components originate far more closely to the
central black hole than the constant line components. A
gravitational redshift effect in this variable component might
thus be present. We measured the differential shifts $\Delta v = \Delta z~c$
of the
strongest rms emission profiles with respect to their mean line
profiles (Fig.~1, Table~1). The observed shifts of the rms profiles
are identified as gravitational redshifts. The mean profiles of the strongest
broad emission lines are not shifted with respect to the
forbidden narrow lines on the other hand. Therefore, the differential shifts
of the rms 
profiles with respect to the narrow lines are identical
to their shifts with respect of the mean profiles.

We calculated the
central black hole mass $M_{\rm grav}$ in  Mrk\,110 using the formula 
(e.g. Zheng \& Sulentic \cite{zheng90})
\[ M_{\rm grav} = c^{2} G^{-1} R \Delta z  .\]

The distances $R$ ($R = c \tau$) of the individual line emitting regions are
known from the cross-correlation analysis (Table 1). We determined
a black hole mass
$M_{\rm grav}$ of $12\pm4$ to $17\pm4\times 10^{7} M_{\odot}$
with a mean value of
\[ M_{\rm grav} = 14\pm3\times 10^{7} M_{\odot} . \]
from the line shifts.
The rms line profile shifts are shown in Fig.~3 as a function of
distance to the center.

The internal shifts of all four emission lines
(H$\alpha$,
H$\beta$, HeI$\lambda$5876, and  HeII$\lambda$4686) 
are different. They depend on the distance of the line
emitting region from the center.
It is intriguing that the black hole masses
we derived from the four different line shifts are identical within
the error limits. This verifies our interpretation of the observed differential
line shifts as gravitaional redshifts.
\begin{figure}
\includegraphics[width=105mm,angle=270]{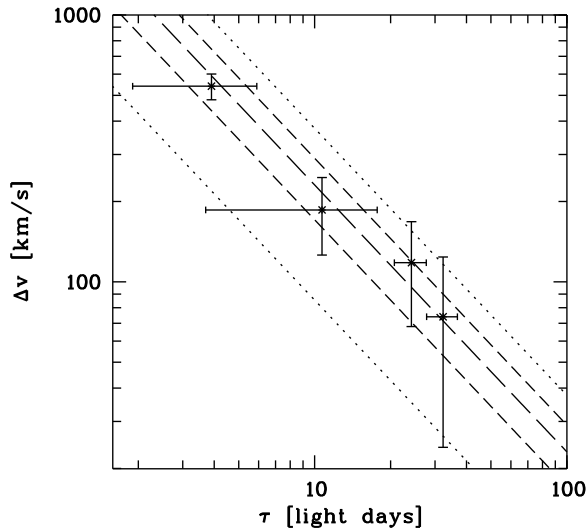}
\vspace{-30mm}
\caption{Redshift of the rms profiles of the HeII$\lambda 4686$,
HeI$\lambda 5876$, H$\beta$ and H$\alpha$ lines as a function of the distance
of the rms line emitting region (see Table~1). The dotted and dashed
curves show computed
lines of gravitational redshifts for central masses of 
$5.0, 10.0, 13.5, 17.0, 22.0\times 10^7 M_{\odot}$ (from bottom to top).}
\end{figure}

The dotted and dashed lines represent theoretical models. They show
the correlation between distance (i.e. cross-correlation lag $\tau$) and
gravitational redshift $\Delta v$ for different central black hole
masses $M_{\rm grav}$ of
5.0, 10.0, 13.5, 17.0, and $22.0\times 10^7 M_{\odot}$,
respectively. 

Other evidence for the
detection of a gravitational redshift has been shown before
in the spectra of the Seyfert galaxy NGC4593 (Kollatschny
et al. \cite{kollatschny97}). In
this galaxy the Balmer lines originate at distances of 2 to 4 light days only.
On the other hand, the quality of the spectra in the NGC4593 variability
campaign was not as good as in Mrk110.  The derived black hole mass
$M_{\rm grav} = 2.3\times 10^{7} M_{\odot}$ in NGC4593 was by a factor of 1.6
higher in comparison to the mass  $M_{\rm orbital}$ derived from the
line widths. An additional hint for a gravitational redshift can be
seen in the HST spectra of the NGC5548 variability campaign. The line
center of the rms CIV$\lambda$1550 profile is redshifted with respect
to the mean profile by about 10\AA\ (see Fig. 4 in Peterson
\cite{peterson01}).
\subsection{Inclination angle of central accretion disk}
The determination of black hole masses in AGNs through their
gravitational redshift effect has clear advantages. The derived
central black hole mass $M_{\rm grav}$ is not affected by the orientation of
the central accretion disk in contrast to the mass $M_{\rm orbital}$
derived from the emission line widths. It has been mentioned before
that only an accretion disk model of the BLR in Mrk\,110 can reproduce
the observed line profile variations (Kollatschny et al. \cite{kollatschny01},
Kollatschny \cite{kollatschny03}).
The mutual
comparison of both mass estimates gives us information on the
orientation angle $i$ of the accretion
disk: $M_{\rm orbital}/M_{\rm grav} = \sin^{2} i$.
For the narrow line Seyfert galaxy Mrk\,110 we derive a nearly pole-on
line-of-sight of the central accretion disk. The inclination angle $i$ amounts
to $21\pm5 \deg$. A disk inclination angle of  $30\pm20 \deg$ has been
previously estimated from line profile variations (Kollatschny
\cite{kollatschny03}). This confirms independently the nearly pole-on view of
the accretion disk in Mrk\,110.  

 The variable HeII line component originates at
a mean distance of 3.9 light-days
($\stackrel{\scriptscriptstyle\wedge}{=}9.8\times 10^{15}$ cm)
only. This distance
corresponds to 230 Schwarzschild radii $r_{\rm s}$
($r_{\rm s}= 2GM_{\rm grav}/c^{2}$) for a
central black hole mass of $14\pm3\times 10^{7} M_{\odot}$. It is very
plausible that the rotation axis of the inner accretion disk is
oriented parallel to the spin axis of the central black hole.
Altogether,
the black hole mass $M_{\rm grav}$  derived from the gravitational redshift
is not only more reliable: a comparison with the
black hole mass $M_{\rm orbital}$
deduced
from the line widths opens up the possibility to determine  the
orientation of the central black holes in radio-quiet AGN. A good test
of the method proposed here would be to check the inclination of optical
broad emission line regions with that of radio/optical jets in
radio-loud objects.

\begin{acknowledgements}
      WK thanks the UT Astronomy Department for warm hospitality during
      his visit. He thanks K. Bischoff and M. Zetzl
      for valuable comments.
      Part of this work has been supported by the
      \emph{Deut\-sche For\-schungs\-ge\-mein\-schaft, DFG\/} grant
      KO 857/24 and DAAD.
\end{acknowledgements}
\end{document}